# The Logic of Strategic Assets: From Oil to AI

Jeffrey Ding and Allan Dafoe[*]
Forthcoming at *Security Studies* (2021)

## Abstract

What resources and technologies are strategic? This question is often the focus of policy and theoretical debates, since the "strategic" designation yields valuable resources and elevated attention. These conversations, however, are frustrated by the ambiguity of the very concept. We offer a theory of when decision-makers should designate particular assets as strategic based on the presence of *important rivalrous externalities* for which socially optimal behavior will not be produced alone by firms or military organizations. We distill three forms of these externalities, which involve *cumulative-*, *infrastructure-*, and *dependency-strategic logics*. While our framework cannot resolve debates about strategic assets, it provides a theoretically grounded conceptual vocabulary to make these debates more productive. To illustrate the analytic value of our framework for thinking about strategic technologies, we examine the U.S.-Japan technology rivalry in the late 1980s, as well as current policy discussions about artificial intelligence.

[*] Jeffrey Ding is a PhD candidate in the Department of Politics and International Relations, University of Oxford, and a researcher at the Center for the Governance of Artificial Intelligence (GovAI). He is currently a predoctoral fellow at the Center for International Security and Cooperation at Stanford University, sponsored by Stanford's Institute for Human-Centered Artificial Intelligence. Allan Dafoe is an Associate Professor at the University of Oxford and the Director of GovAI, at the Future of Humanity Institute. Correspondence email: jding99@gmail.com

For useful feedback we thank: Emefa Agawu, Amanda Askell, Markus Anderljung, Miles Brundage, Ben Buchanan, Alexis Carlier, Richard Danzig, Max Daniel, Alexandre Debs, Ben Garfinkel, Ozzie Gooen, Jade Leung, Theodore Moran, Nuno Monteiro, Luke Muehlhauser, Toby Shevlane, Remco Zwetsloot, John Zysman, the anonymous reviewers, participants at the University of Oxford International Relations Research Colloquium and at seminars at the Centre for the Governance of AI, housed at the Future of Humanity Institute, and especially Duncan Snidal and Steven Weber.

**Overview**

In March 2018, when the Office of the U.S. Trade Representative released its Section 301 report on China's unfair trade practices — one of the first volleys in a trade war — astute observers noted that the report singled out one Chinese technology plan: "Made in China 2025 [*zhongguo zhizao 2025*]." Amidst the bluster of tariffs on steel and soybeans, these analysts understood that "Made in China 2025," which prioritized ten "strategic industries (*zhanlüe chanye*)," posed the "real existential threat to U.S. technological leadership."[1] Indeed, competition over "strategic" goods and technologies has become a focal point in the broader U.S-Sino rivalry, though the crucial step of clarifying what makes an asset "strategic" has often been neglected.[2]

This definitional vagueness is neither new nor limited to the U.S.-China rivalry. For centuries policymakers and theorists have debated which goods and technologies deserve the "strategic" descriptor. These conversations matter. David A. Baldwin's characterization of the debates over strategic assets, written in 1985, remains true today: "Widespread misunderstanding of the concept of 'strategic goods' is one of the biggest impediments to intelligent discussions of economic statecraft."[3] However, even as nations are increasingly concerned about strategies to build up technological advantages over their rivals, much more work needs to be done to understand the underlying logic of what makes an asset strategic.

---

[1] There were 116 mentions of "Made in China 2025" in the Section 301 report. The plan was first issued by the Chinese State Council in May 2015. The ten strategic sectors included high-end numerical control machinery and robotics, energy-saving and new energy vehicles, biopharmaceuticals and high-performance medical devices, etc. Lorand Laskai, "Why Does Everyone Hate Made in China 2025?" *Council on Foreign Relations*, 28 March 2018, https://www.cfr.org/blog/why-does-everyone-hate-made-china-2025

[2] For instance, in October 2018 the U.S. Department of Treasury established a pilot program that increased scrutiny of inward foreign investment in "critical technologies," a broad category that includes defense equipment as well as a similarly broad subcategory of "emerging and foundational technologies" which has not been defined. U.S. Department of the Treasury, "Determination and Temporary Provisions Pertaining to a Pilot Program To Review Certain Transactions Involving Foreign Persons and Critical Technologies," 11 October 2018, https://home.treasury.gov/system/files/206/FR-2018-22182_1786904.pdf

[3] David A. Baldwin, *Economic Statecraft* (Princeton, NJ: Princeton University Press, 1985), 214.

How should national leaders identify strategic assets? In this paper, we present a unified theoretical framework based on an asset's connection to *important rivalrous externalities*, such that optimal transactions involving these assets will not be achieved by markets and individual national security entities themselves. Strategic assets are those for which attention from the highest levels of the state is required to secure national welfare against interstate competition.

Our theory of strategic assets offers a conceptual framework for clarifying policy debates over technology strategy. Mirroring George and Smoke's characterization of the policy function of deterrence theory, our theoretical work can best be understood as serving a "diagnostic function," providing "assistance to policymakers in assessing the configuration of a situation."[4] Instead of how to deter other states, our focus is on how to identify strategic assets.

This framework is roughly captured in the following "strategic formula" for goods and technologies:

**Strategic Level of Asset = Importance * Externality * Nationalization**

The strategic level of an asset is a product of the following three factors:

1. *Importance*: an asset's economic and/or military utility (some sectors, e.g. freight transport, contribute more to economic growth than others, e.g. high-end fashion).
2. *Externality*: the economic and/or security externalities associated with an asset, such that uncoordinated firms and individual military organizations will not optimally attend to the asset (e.g. the positive externalities generated by research into foundational technologies, which private actors under-invest in because they do not capture all the spillover benefits).[5]

---

[4] Alexander L. George and Richard Smoke, "Deterrence and Foreign Policy," *World Politics* 41, no. 2 (January 1989): 170–82, 180; Alexander L. George and Richard Smoke, *Deterrence in American Foreign Policy: Theory and Practice* (New York, NY: Columbia University Press, 1974), 616-642. Part two provides a longer explication of our method.
[5] Even the most important assets, such as nuclear weapons, may score low in this factor if states have already internalized all the externalities associated with these assets.

3. *Nationalization*: the degree to which these externalities are rivalrous between nations. Some assets, such as fundamental research in medicine, generate positive externalities that may easily diffuse to other rival nations, which limits their strategic level.[6]

Of these three factors in our framework, we will focus on what we regard as the most illuminating aspect of this equation: the existence and character of externalities that demand the attention of the state. Typically applied to the behavior of private firms, externalities can also pertain to the actions of military entities, such as the Navy, which have an incentive structure that does not wholly internalize the interests of other sub-national actors. Basing the framework on externalities also roots it in existing scholarship at the intersection of economics and national security.[7]

Externalities come in many shapes, but we distill three forms of these externalities — the *cumulative-*, *infrastructure-*, and *dependency-*strategic logics — that cover a substantial range of the strategic qualities of assets (Table 1). The **cumulative-strategic logic** involves assets and sectors with high barriers to entry linked to cumulative processes, such as first-mover dynamics, incumbency advantages, and economies of scale. These high barriers to entry lead the market to under-invest, and military organizations to require explicit state support to achieve nationally optimal investments. Aircraft engines [1945-present] serve as a representative example. Even with government support, China's defense firms still lag behind the top producers of aircraft engines due to high R&D costs and steep learning curves.[8]

The **infrastructure-strategic logic** involves assets that generate positive spillovers across the national economy or military system, which sub-national actors (e.g. firms or military branches) under-invest in because they do not appropriate all the associated gains. These are often foundational technologies that complement and upgrade the national technological system. A representative example is railroads [1850-1890], which generated enormous positive spillovers for the U.S. economy by enabling new patterns of labor mobility, economies of scale for manufacturing, and wholesale food distribution channels.[9]

---

[6] Nevertheless, such assets may still be worthy of government attention, as they provide absolute benefits.
[7] William J Norris, *Chinese Economic Statecraft: Commercial Actors, Grand Strategy, and State Control* (Ithaca, NY: Cornell University Press, 2016); Andrew Kennedy and Darren J. Lim, "The Innovation Imperative: Technology and US–China Rivalry in the Twenty-first Century," *International Affairs* 94, no. 3 (2018): 553-572.
[8] Stephen G. Brooks and William C. Wohlforth, "The Rise and Fall of the Great Powers in the Twenty-first Century: China's Rise and the Fate of America's Global Position," *International Security* 40, no. 3 (2016): pp. 7-53, 38.

The ***dependency-strategic logic*** involves assets that are not supplied by a robustly open and competitive market, making them vulnerable to cutoffs. This often arises from physical, organizational, or national concentration in the supply chain, such that an adversary could plausibly intervene to reduce supply. These assets also must be essential, and thus have few substitutes. Nitrates [1914-1918] are a representative example, as demonstrated by the British naval blockade's effect on Germany's supply chain for explosives, preventing nitrate imports from Chile, the world's principal supplier.

| TABLE 1 The Logics | Description | Examples (Economic) | Examples (Military) | Examples (Both) |
|---|---|---|---|---|
| **Cumulative-strategic** | Have high barriers to entry, due to first-mover dynamics, incumbency advantages, economies of scale, or other cumulative dynamics. | Digital social networks [2000-present][10] | Stealth fighters [1945-present] | Aircraft engines [1945-present] |
| **Infrastructure-strategic** | Generates (diffuse) positive spillovers across the national economy or military system. These are often fundamental technologies that upgrade the national technological system. | Electricity [1890-1920] | Radar [1930-1945] | Railroads [1850-1890] |
| **Dependency-strategic** | Supply characterized by extra-market dynamics and few substitutes. | Platinum [1960-present] | Nitrate [1914-1918] | Integrated circuits [1980-present] |

These logics illustrate that the strategic level of an asset is not intrinsic to the good or technology itself. Not only is an asset's strategic level shaped by features of the international environment (e.g. the rate of cross-border diffusion of technology) but it is also affected by the particular strategy pursued by a state (e.g. one oriented around a land army, versus navy, versus economic might, versus soft power). Motivated by the present environment of U.S.-Sino rivalry, we focus on the strategy of a great power concerned about growing its economic and military strength vis-à-vis that of its peer competitors;[11] however, we also emphasize that our framework is valid and useful for contexts in which states are pursuing different strategies.[12]

---

[9] Dave Donaldson and Richard Hornbeck, "Railroads and American Economic Growth: A 'Market Access' Approach," *Quarterly Journal of Economics* 131 no. 2 (2016): 799-858.
[10] This notation gives our rough gauge of the date range for which the asset remained at a high-level of strategic significance for industrial great powers.

Bringing together scattered notions of strategic goods and technologies from international political economy and security studies, our framework contributes to the literature on the political economy of national security.[13] Existing scholarship on grand strategy rightly emphasizes the growing significance of technological instruments.[14] While these studies shed insight into the relevant actors, doctrines, and consequences of various strategies, what is often missing in these discussions is the objects themselves — that is, how to determine which assets should be the target of statecraft.

Specifically, the designation of strategic assets drives at concerns about trading with the enemy, which is central to theories about economic interdependence and conflict.[15] However, these ongoing debates are inhibited by "the analytical problem of how to distinguish strategic goods from others, particularly when the meaning of 'strategic' may vary across time and space."[16] Our framework provides a clearer conception of strategic goods by disentangling their military and/or economic utility (e.g. the importance of having foodstuffs or weapons in a war effort) from the externalities associated with their production (e.g. the extent to which supply of foodstuffs or weapons is concentrated in a particular country).

---

[11] For strategies pursued by nonmajor powers, see: William I. Hitchcock, Melvyn P. Leffler, and Jeffrey W. Legro, *Shaper Nations: Strategies for a Changing World* (Cambridge, MA: Harvard University Press, 2016); Thierry Balzacq, Peter Dombrowski, and Simon Reich, eds., *Comparative Grand Strategy: A Framework and Cases* (New York, NY: Oxford University Press, 2019).
[12] Suppose, for instance, a state wants to base its strategy on soft power. Our framework gives an initial cut for where strategic assets may be positioned: sources of cultural capital dominated by a single country with high barriers to entry, such as the Hollywood film industry (cumulative-strategic); instruments, such as strict national standards against corruption, that prevent companies from undermining a nation's brand (infrastructure-strategic); social media sites that dominate coverage for a particular country (dependency-strategic). Separately, a state's strategy may not give equal weight to economic and military might. For our framework, this would affect the "importance" of a particular asset but not the associated externality. We thank an anonymous reviewer for pointing this out.
[13] Michael Mastanduno, "Economic Statecraft, Interdependence, and National Security: Agendas for Research," *Security Studies* 9, no. 1–2 (1999): 288–316. Jonathan Kirshner, "Political Economy in Security Studies after the Cold War," *Review of International Political Economy* 5, no. 1(1998): 64-91. Jonathan D. Caverley, "United States Hegemony and the New Economics of Defense," *Security Studies* 16, no. 4 (2007): 598–614..
[14] Kennedy and Lim, "The Innovation Imperative"; Mark Z. Taylor, "Toward an International Relations Theory of National Innovation Rates," *Security Studies* 21, no. 1 (2012): 113–52; Elizabeth Thurbon and Linda Weiss, "Economic Statecraft at the Frontier: Korea's Drive for Intelligent Robotics," *Review of International Political Economy* (2019), 1–25.
[15] Jack S. Levy and Katherine Barbieri, "Trading with the Enemy During Wartime," *Security Studies* 13, no. 3 (2004): 1–47; Peter Liberman, "Trading with the Enemy: Security and Relative Economic Gains," *International Security* 21, no. 1 (1996):

We also contribute to broader conversations over strategy and statecraft targeted at specific technologies, which increasingly extend beyond controlling trade.[17] In many cases, export restrictions on strategic assets hamper a nation's innovation system from "running faster," tradeoffs between two logics captured by our framework.[18] By targeting externalities generated by certain assets, our framework can help craft a more multidimensional and pragmatic technology strategy. This is in line with fifty years of research on scientific and technological competitiveness, which attributes success to a country's general commitment to solving market failures rather than to the role of any particular institution or doctrine.[19]

The paper proceeds in four sections. The first part reviews the literature on strategic goods and technologies, revealing the myriad and oft-confused understandings of the concept. We synthesize from this literature our three underlying logics of strategic assets: cumulative-strategic,

---

147–75; Norrin M. Ripsman and Jean-Marc F. Blanchard, "Commercial liberalism under fire: Evidence from 1914 and 1936," *Security Studies* 6, no. 2 (1996): 4-50.
[16] Levy and Barbieri, "Trading with the Enemy During Wartime," 11.
[17] Beverly Crawford, *Economic Vulnerability in International Relations: the Case of East-West Trade, Investment, and Finance* (New York, NY: Columbia University Press, 1993; Glenn Fong, "Breaking New Ground or Breaking the Rules," *International Security* 25, no. 2 (2000): 152-186; Nuno P. Monteiro, *Theory of Unipolar Politics* (Cambridge, U.K.: Cambridge University Press, 2014), 124-142; Brooks and Wohlforth, "The Rise and Fall of the Great Powers in the Twenty-first Century."
[18] Hugo, Meijer, *Trading with the Enemy: The Making of US Export Control Policy toward the People's Republic of China* (Oxford, UK: Oxford University Press, 2016).
[19] Mark Z. Taylor, *The Politics of Innovation: Why Some Countries Are Better Than Others at Science and Technology* (Oxford, UK: Oxford University Press, 2016), 277. For an analysis of why the U.S. should adopt a more pragmatic approach to policy-making, which articulates the logic behind policy yet eschews the rigidity of strategizing, see David M. Edelstein and Ronald R. Krebs, "Delusions of Grand Strategy: The Problem With Washington's Planning Obsession," *Foreign Affairs* 94, no. 6 (2015): 109-116.

infrastructure-strategic, and dependency-strategic. We fill out our externality-based framework of strategic assets in part two, where we explain how firms and militaries fail to adequately internalize the benefits or risks of a particular good or technology. Part three illustrates the analytical value of our framework by examining the U.S.-Japan technology rivalry in the 1980s and 1990s. Specifically, we highlight consistent missteps in U.S. efforts to identify strategic assets, which failed to achieve purported goals. Finally, part 4 applies our framework to artificial intelligence (AI), which has become central to current discussions about international technological competition.

## Part I: Evolution of an Idea

In 1985, Baldwin wrote in his authoritative text on economic statecraft, "The controversy over what constitutes a 'strategic good' has been going on for thirty years."[20] Over thirty years later, scholars and policymakers still struggle with this ambiguous concept, leading many to give up on the exercise altogether and rely on gut feel — "they know a strategic industry when they see one."[21] Previous theorizing about strategic assets can be grouped into three camps, distinguished by whether the focus of analysis is on: 1) military significance, 2) substitutability, or 3) strategic trade. Taking the perspective of a strategist concerned with the national interest, our framework for strategic assets highlights gaps and integrates insights from each of these camps.

### *Military Significance Camp*

In the first camp, scholars emphasize the "military significance" of certain assets, arguing that the strategic quality of goods and technologies is determined by their military utility.[22] The underlying assumption is that goods and technologies are "only strategic if they can be used for war, or converted for war, or processed into war-type goods."[23]

This view of strategic assets is prevalent in export control and defense industrial policy in countries around the world, especially the United States. For example, from 1989 to 1992, the U.S. Department of Defense (DoD) published annual critical technology plans, which designated twenty technologies as critical for the long-term qualitative superiority of U.S. weapons systems.[24] The U.S. export regime is based on a conception of strategic assets tied to military end uses and users.[25]

---

[20] Baldwin, *Economic Statecraft*, 106. Baldwin traces this debate back to Yuan-Li Wu, *Economic Warfare* (Upper Saddle River, NJ: Prentice-Hall, 1952).
[21] David J. Teece, "Support Policies for Strategic Industries: Impact on Home Economies," in *Strategic Industries in a Global Economy: Policy Issues for the 1990s*, ed. OECD (Paris: CEDEX, 1991), 35-60, 36.
[22] See, for example, Gunnar, Adler-Karlsson, *Western Economic Warfare* (Stockholm, SE: Almqvist & Wiksell, 1968), 3.
[23] Thomas Schelling, 1958, *International Economics* (Boston, MA: Allyn and Bacon, Inc.), 500.
[24] Crawford, "Economic Vulnerability in International Relations," 16-17.
[25] For an analysis of the challenges globalization and dual-use technologies pose for U.S. export control policy, see Meijer, *Trading with the Enemy*.

From the perspective of our framework, military assets often are strategic because they are often important *and* often exhibit one of the three strategic logics. However, notably, many military assets fail one of these criteria, and thus by our framework should not be regarded as strategic. Some important military assets are readily supplied through global markets, or produced domestically through existing organizational capacity, and thus do not require the high-level attention of the state because they do not exhibit rivalrous externalities. In the context of modern militaries, various types of missiles, machine guns, and other small arms are examples of assets which are militarily significant but not strategic under our framework.[26]

*Substitutability Camp*

The substitutability camp's view of strategic assets loosely corresponds to our framework's third logic (*dependency-strategic*). These thinkers base a good's strategic level on its substitutability, as captured by the degree to which it is critical to a significant economic or military process as well as the availability of substitutes for the good. In line with the dependency-strategic logic, Osgood defines a strategic good as "an item for which marginal elasticity of demand is very low and for which there is no readily available substitute."[27]

Elaborating this substitutability logic, Ripsman and Blanchard's four-part "strategic goods test" is an exemplary, rigorous attempt to differentiate among classes of goods and technologies in Anglo-German economic relations before the First World War.[28] They first evaluate which goods were essential to national defense and economic well-being (*importance* in our equation) and then assess the impact of a supply cut-off by analyzing whether substitutes could have alleviated any disruptions (a particular kind of *externality* in our equation).

---

[26] Using machine guns as an example, we expand on why military significance is insufficient for the designation of an asset as "strategic" in part two. For evidence that U.S. competitors have been able to catch-up in missiles, due to muted cumulative effects, see: Brooks and Wohlforth, "The Rise and Fall of the Great Powers in the Twenty-first Century," 38.
[27] Theodore K. Osgood, "East-West Trade Controls and Economic Warfare" (Ph.D. diss., Yale University, 1957), 89; Baldwin, *Economic Statecraft*, 215.
[28] Ripsman and Blanchard, "Commercial Liberalism under Fire."

Primarily concerned about dependency risks, the substitutability camp does not look at other externalities that can give rise to strategic assets. Furthermore, the condition that the externality be rivalrous is rarely made explicit, plausibly because supply dependence is usually political, and thus can be made rivalrous. However, some dependence is in principle not rivalrous, such as two belligerents for whom a natural disaster disrupts their supply of oil, impacting them roughly equally.

*Strategic Trade Camp*

A third school of thought, the "strategic trade" camp, often lumps together the two strategic logics missed by the substitutability camp. Strategic trade theorists highlight the extent to which particular industries confer large first-mover advantages, present high barriers to entry, and/or yield enormous spillovers.[29] These considerations have risen in prominence alongside the liberalization of foreign direct investment flows, which enabled the consolidation of large-scale oligopolies.[30]

Strategic traders aim to ensure that their national economy can both compete in industries with high learning curves and benefit from spillovers associated with the production of certain assets. Still, analysis from the strategic trade camp is largely limited to the economic domain and rarely differentiates between infrastructural and cumulative externalities; our framework rectifies both of these shortcomings.

---

[29] Paul Krugman, ed., *Strategic Trade Policy and the New International Economics* (Cambridge, MA: MIT Press, 1986).
[30] Wolfgang Michalski, "Support Policies for Strategic Industries: An Introduction to the Main Issues," in *Strategic Industries in a Global Economy: Policy Issues for the 1990s*, ed. OECD (Paris: CEDEX, 1991), 7-15, 8.

## Part II: Conceptual Framework

Under our framework, strategic assets are those for which there is an *externality* that is both *important* and *rivalrous*. As captured by the strategic formula in the overview, the strategic level of an asset is a product of these three factors, not a sum of three addends. Thus, these three criteria are each necessary and jointly sufficient for an asset to qualify as strategic (Table 2).[31]

| Table 2: Tri-Logic Framework for Strategic Assets<br>I = importance, E = externality, N = extent to which externality differentially accrues to one nation vs rival ones.<br>1s and 0s are a binary simplification. All three are continuous variables. | | | |
|---|---|---|---|
| | I * E * N | I * E * N | **I * E * N** |
| **Strategic Logic** | 1, 0, 0 | 1, 1, 0 | **1, 1, 1** |
| Cumulative | Steel [1990s-present][32] | ITER [1985-present][33] | **Aircraft engines [1945-present]** |
| Infrastructure | Real estate | Publications in basic science | **Recombinant DNA tech [1980-present][34]** |
| Dependency | Wheat | Ozone | **Integrated circuits [1980-present]** |

Consider two examples in the cumulative-strategic domain. First, assets that present externalities but have low levels of "importance" are not strategic. A new technology for brewing that exhibits strong "learning by doing" characteristics may generate barriers to entry, but the scale of the brewing industry does not have a substantial effect on the economic or military power of nations. Conversely, an asset can be extremely important but not strategic. M240 machine guns are essential equipment for infantry platoons and armored vehicles, but they are so easy for most nations to build or acquire that they do not exhibit strategic externalities.

---

[31] The concept of an asset's strategic level is a continuous one, as all three factors are continuous. Each condition must be present to a sufficient extent for the asset to have a high strategic level. Formally, this can be expressed as a product of the three conditions, or as these conditions being continuous versions of necessary and sufficient conditions. Geortz 2006, 42. Our focus is on analyzing the positive and negative poles of the concept of strategic assets, though we acknowledge that there are gray zones where the cutoffs for the three criteria are unclear, as is the case with more theorized concepts, such as whether Switzerland is a corporatist or noncorporatist system, or whether Malawi qualifies as a democracy. We thank an anonymous reviewer for bringing up this point.
[32] Part three explains why steel became less strategic.
[33] ITER, originally called the International Thermonuclear Experimental Reactor, is a 35-nation project to demonstrate the large-scale feasibility of fusion. Any competitor project, whether a single-firm or multinational effort, faces enormous barriers to entry, but any accumulated gains from ITER will disperse across countries.
[34] For an empirical demonstration that recombinant DNA technology is a general-purpose technology, with the potential to improve productivity across many sectors, see: Maryann P. Feldman and Ji Woong Yoon, "An Empirical Test for General Purpose Technology: An Examination of the Cohen–Boyer rDNA Technology," *Industrial and Corporate Change* 21, no. 2 (2011): 249-275.

This section further explores how these three strategic logics function in the economic and military domain. For each logic, we describe the mechanics of the externality, provide examples of strategic assets, and differentiate our interpretation of the logic from related, influential concepts (cumulative to Van Evera's "cumulative resources"; infrastructure to dual-use; dependency to Hirschman's "dependence"). We conclude by highlighting the possible interactions between multiple logics, including scenarios in which assets are linked to multiple types of externalities or multiple logics come into conflict with each other.

*Strategic Logic #1: Cumulative*

The cumulative-strategic logic is underpinned by cumulative processes that entrench barriers to entry, a broad concept which covers long investment timelines, first-mover advantages, winner-take-all dynamics, learning-by-doing, etc. The potency of the cumulative effect can vary: a winner-take-all phenomenon, fueled by strong network effects, constitutes a strong version of the logic, while modest returns to scale generated by "learning by doing" evince a weak version of the logic.

The cumulative-strategic logic is relatively well-understood in the economic domain. Most markets do not yield substantial rents because interfirm competition moves the surplus to consumers. However, competition is weakened in cases of production characterized by cumulative effects, such as when returns only accrue after long time scales or risky bets, under strong economies of scale, and in the presence of other first mover advantages.[35] Industries often identified as cumulative-strategic include semiconductors, commercial aircraft, and telecommunications.[36] As a byproduct of these cumulative processes, these strategic assets generate rents that accrue to the firms who were able to overcome the barriers.

---

[35] On the subject of how the intervention of foreign governments affects the strategic calculus for cumulative gains in certain industries, see Marc L. Busch, *Trade Warriors: States, Firms, and Strategic-Trade Policy in High-Technology Competition* (Cambridge, U.K.: Cambridge University Press, 2001).

Certain types of defense technology also exhibit the cumulative-strategic logic. For example, prime contractors (mostly based in the U.S.) benefit from network effects that come from controlling the systems integration of the arms production industry. Comparing systems integration technologies to "killer applications" and dominant standards such as Microsoft's Windows operating system, Caverley highlights how customers/suppliers have an incentive to participate in weapons programs backed by U.S. prime contractors since each new customer/supplier enhances the weapon's value for everyone in the network.[37]

Cumulative dynamics associated with "learning by doing" manifest in the interaction between some military technologies and the organizations that produce them. For instance, A. Gilli and M. Gilli argue that the required technical knowledge to make stealth fighters has become increasingly organizational in nature, thereby limiting their diffusion.[38] Since this organizational knowledge has accumulated in the collective memory of the U.S. military, rival nations cannot acquire it through licenses, stealing blueprints, or even kidnapping engineers.

Lastly, it is important to relate our "cumulative-strategic" concept to Van Evera's conceptualization of "cumulative resources."[39] Van Evera defines a cumulative resource as one that "helps its possessor to protect or acquire other resources,"[40] and this concept has inspired research on the cumulativity of territory and conquest.[41] He specifies that the cumulativity of a resource is a function of the utility of the resource for acquiring or protecting other resources as well as the cost of extracting the resource from its territory.[42]

---

[36] Helen V. Milner and David B. Yoffie, "Between Free Trade and Protectionism: Strategic Trade Policy and a Theory of Corporate Trade Demands," *International Organization* 43, no. 2 (1989): 239-272.

[37] Caverley, "United States Hegemony and the New Economics of Defense," 605-607.
[38] Andrea Gilli and Mauro Gilli, "Why China Has Not Caught Up Yet: Military-Technological Superiority and the Limits of Imitation, Reverse Engineering, and Cyber Espionage," *International Security* 43, no. 3 (2019): 141-189, 162-163.
[39] We thank Nuno Monteiro for pointing us to Van Evera's concept of cumulative resources. Stephen Van Evera, *Causes of War: Power and the Roots of Conflict* (Ithaca, New York: Cornell University Press, 1999), 105-116.
[40] Ibid., 105.

Under our framework, Van Evera's conception of cumulativity primarily factors into an asset's *importance,* whereas our notion of cumulativity (cumulative-strategic) functions as an externality. For instance, Van Evera asserts that uranium ore became more cumulative after the advent of nuclear weapons.[43] Our framework does not preclude the valuation of an asset's potential to enable the acquisition of other resources. However, the fungibility of resources means that most assets help their possessor protect another resource and thus qualify as cumulative under Van Evera's conception. We do view uranium ore as becoming more important in the nuclear age, but we do not consider it to be cumulative-strategic as one nation's investment in uranium (short of trying to corner the global market) did not lead to barriers to entry for other countries to acquire uranium.[44]

*Strategic Logic #2: Infrastructure*

The second logic, which we call "infrastructure-strategic," involves assets that generate large positive spillovers that cannot be internalized by the initial innovators. These assets typically upgrade the national technological system, thereby benefiting other firms in the same industry or related industries.[45] In the context of this logic, the *rivalrous* variable measures the degree to which these spillovers are largely contained within national borders or amongst allies (such as transportation networks), as opposed to being global (such as say basic advances in medicine). Many technical advances, especially with increasingly rapid global diffusion, have global impacts, pushing out the technological frontier to the benefit of all parties.[46] However, even in a world of increased cross-border diffusion of technology, the spillovers from many infrastructure-strategic innovations cluster geographically, differentially advantaging some nations over others.[47]

---

[41] Stephen G. Brooks, *Producing Security* (Princeton, NJ: Princeton University Press, 2005); Richard Rosecrance, *The Rise of the Virtual State* (New York, NY: Basic Books, 2000).
[42] Van Evera, *Causes of War*, 106.
[43] Ibid., 107.
[44] The dependency-strategic logic may have applied to uranium for some countries, but over time it was discovered that uranium ore was plentiful and relatively widespread across countries. Jonathan E. Helmreich, *Gathering Rare Ores: The Diplomacy of Uranium Acquisition, 1943-1956* (Princeton, NJ: Princeton University Press, 1986)
[45] Michalski, "Support Policies for Strategic Industries."

Since this logic operates through interconnections that may benefit both the economic and military realm, infrastructure-strategic assets are often characterized as "dual-use." Most of the time, economists are not trained to consider externalities in the national security domain, and military strategists do not focus on the effects of military technologies in the economic realm. Of the three strategic logics we highlight, the infrastructure-strategic logic is most helpful in illustrating the value of a framework that accounts for the strategic qualities of assets across both the economic and military domains.

Indeed, many dual-use technologies (e.g. computers) can be considered infrastructure-strategic, as actors oriented toward maximizing benefits in either the security or economic domain do not internalize the cross-domain spillovers.[48] One particularly notable instance of spillover was the U.S. military's investment in ARPAnet to secure the flow of information in the event of a nuclear attack, which stimulated development of technologies critical for the development of the internet.[49] Flowing in the opposite direction, spin-ons from the commercial sector to military applications have now become more important than spin-offs in the reverse direction.

---

[46] We thank Theodore H. Moran for this point. Whether technological knowledge spillovers are global or local has been much debated in the empirical economics literature. One analysis has shown that while local spillovers are still important, the extent to which knowledge spillovers decline with distance has fallen by 20 percent, partly due to the increase in foreign R&D by technology producers. Wolfgang Keller, "Geographic Localization of International Technology Diffusion," *American Economic Review* 92, no. 1(2002): 120-142.
[47] Michael Borrus,, Laura D'Andrea Tyson, and John Zysman, "Creating Advantage: How Government Policies Shape International Trade in the Semiconductor Industry," in *Strategic Trade Policy and the New International Economics*, ed. Krugman, 99-114, 94. See also Busch, *Trade Warriors*.
[48] Michael L. Dertouzos, Robert M. Solow, and Richard K. Lester, *Made in America: Regaining the Productive Age* (Cambridge, MA: MIT Press, 1989), 115.
[49] U.S. military planners did not initially view ARPAnet as an infrastructure-strategic technology with possible spillover effects, as the internet was more a product of happenstance than intentional strategy.

Advancements in fiber optics, for instance, play an important role in the modern economy by transmitting data at high speeds, but they also have spin-on effects for national security by improving missile guidance capabilities.[50] In other domains, such as civilian-aircraft and military-aircraft technology, the linkages between some civilian and military assets are becoming more tenuous.[51] Thus, evaluating the dual-use potential of an asset requires understanding how the connection between civilian and military assets is evolving.

In the economic domain, infrastructure-strategic assets are often foundational technologies that transform the outputs and production processes of a wide range of industrial sectors, and thus exert a profound effect on the competitiveness of national economic systems.[52] Railroads, for instance, generated enormous positive spillovers by increasing labor mobility, enabling economies of scale for manufacturing, and expanding transportation of perishable products and natural resources.[53] Additionally, the semiconductor industry is often characterized as providing extremely large spillovers to downstream electronics applications, though some scholars have questioned if private actors actually under-invest in this industry.[54]

In the military domain, the infrastructure-strategic logic characterizes technologies that can upgrade a wide range of military capabilities but are underappreciated due to entrenched organizational interests and lack of coordinated investment. As is often the case with market competition in the economic realm, competition between military services ("The Interservice Model of Military Innovation"), or even within branches of the same military service ("The Intraservice Model of Military Innovation"), efficiently promotes military innovations.[55] However, competition also causes militaries to under-invest in infrastructure-strategic assets that provide benefits across branches and services and often require a common set of technical specifications. For instance, during the late 1980s, though the U.S. DoD recognized the growing centrality of software across many military platforms, bureaucratic obstacles prevented standardization of programming languages and investments in advancing software technologies.[56]

---

[50] Richard J. Samuels, *Rich Nation, Strong Army: National Security and the Technological Transformation of Japan* (Ithaca, NY: Cornell University Press, 1994), 30.
[51] Office of Technology Assessment, *The Defense Technology Base: Introduction and Overview* (Washington, DC: U.S. Government Printing Office, 1988), 30.
[52] Giovanni Dosi, Laura D. Tyson, and John Zysman, "Trade, Technologies, and Development: A Framework for Discussing Japan," in *Politics and Productivity: The Real Story of Why Japan Works*, eds. Chalmers Johnson, Laura D. Tyson, and John Zysman (New York: Harper Business, 1989), 3-38, 8-10.
[53] Donaldson and Hornbeck, "Railroads and American Economic Growth."
[54] For a study that shows semiconductor firms are able to capture most of the spillovers from R&D, see Richard C. and Peter C. Reiss, "Cost-reducing and Demand-creating R&D with Spillovers," *RAND Journal of Economics* 19, no. 4 (1988): 538-556.

The radar [1930-1945] is another illustrative example. Despite having an early lead in developing radars more advanced than the British, Germany failed to realize the radar's potential due to interservice rivalry. The German navy had started work on radar in the early 1930s but did not share any information with the German air force. German leaders also failed to establish a liaison mechanism between the radar units, the fighter units, and the command organization. As a result, when the British conducted a bomber raid on Germany two days after declaring war, German radar detected the bombers, but no fighters were sent out to intercept them.[57] In contrast, the British rapidly integrated radar into a battle-ready air defense system — a process that involved standardizing updates on the number, course, and heading of enemy aircraft — that took full advantage of its infrastructure-strategic attributes.[58]

*Strategic Logic #3: Dependency*

The dependency-strategic logic distills ideas from the substitutability camp into the language of externalities. Our framework highlights relations of dependence that are not internalized by private actors, namely economic transactions involving goods and technologies where a concentration of foreign suppliers imposes a negative externality for the importing state, represented by the potential economic and security costs of being cut off from accessing these items.[59] Individual firms do not fully internalize the downside of a cut-off for the nation's economy or military, for which continued access to these dependency-strategic assets is at risk due to the lack of substitute goods and alternative suppliers. It is important to differentiate dependency-strategic assets from advanced technologies further downstream, such as nuclear weapons and stealth technology, which states do not need to purchase on an ongoing basis.

---

[55] Adam Grissom, "The Future of Military Innovation Studies," *Journal of Strategic Studies* 29, no. 5 (2006): 905–34.
[56] Nance Goldstein, "Institutional Resistance to the Demands of a New Information Technology: Software R&D in the US Defense Department in the 1980s," *International Review of Applied Economics* 7, no. 1 (1993): 26–47.
[57] Azriel Lorber, "Technological Intelligence and the Radar War in World War II," *Royal Canadian Air Force Journal* 5, no. 1 (2016): 52-65, 55-56.
[58] Stephen P. Rosen, "New Ways of War: Understanding Military Innovation," *International Security* 13 (1988), 134-168, 143-149.

Dependency concerns affect both the military and economic domains. At the military end, there are dependency-strategic goods that enable important military functions. For instance, the German military's supply of explosives was severely hampered in the lead-up to World War I due to the British blockade, which cut off German access to nitrate exports from Chile, producers of almost 80 percent of the world's nitrates.[60] Other dependency-strategic assets fuel primarily economic functions. For instance, separate from other strategic stockpiles of materials for military needs, the U.S. Geological Survey maintains a list of minerals critical to economic well-being.[61]

Some goods are critical for both economic and military processes. Oil, the "strategic commodity second to none," is the classic case of such an asset.[62] Before and during World War I, the British set fire to oil fields and the Germans torpedoed freighters, all to prevent the other side from powering their military industries.[63] In the lead up to World War II, U.S. grand strategy engineered conditions in which Japan was dependent on the U.S. for 80 percent of its oil supplies, making the 1941 U.S. oil embargo have devastating effects on Japan's military forces.[64] Additionally, because oil is crucial for industrial economies and consumed in such quantities that it is difficult to stockpile, countries have deployed the "oil weapon," i.e. threatened to or actually cut off oil shipments to other countries, as a tool of economic coercion.[65]

---

[59] Many of these ideas were introduced in Albert Hirschman, *National Power and the Structure of Foreign Trade* (Reprint, Berkeley: University of California Press, [1945] 1980).
[60] Had it not been for Germany's development of synthetic nitrogen, Germany may have lost the war much earlier. William H. McNeil, *The Pursuit of Power: Technology, Armed Force, and Society since AD 1000* (Chicago: IL.: University of Chicago Press, 1982), 79-102.
[61] One of the 23 strategic minerals is platinum, for which 90 percent of production since 1990 has come from two countries: Russia and South Africa. Zientek et al. 2017.
[62] Daniel Yergin, *The Prize: The Epic Quest for Oil, Money, and Power* (New York, NY: Simon & Schuster, 1990), 163.
[63] Ibid., 151-168.

This third strategic logic builds upon Hirschman's concept of dependence, which he defines broadly as "that part of a country's well-being which it is in the power of its trading partners to take away."[66] Hirschman centers his analysis of dependence on the concentration of a country's *aggregate* imports and exports to shed light on bilateral channels of influence. Expanding on this analysis, our framework examines the concentration of suppliers for *specific* goods. These assets highlight not just bilateral power relations but also situations of dependence on suppliers that are politically unstable or vulnerable to natural disasters.

*The Logics in Combination*

While we have largely analyzed these strategic logics in isolation, in any given case these logics can overlap, complementing, attenuating, or complicating each other. States should pay especially close attention to those technologies and goods that exhibit multiple strategic logics. First, if multiple logics are operative, then the asset is, all else equal, more strategic. Second, multiple logics may call for a diverse set of policy responses to address each externality, complicating the policy problem.

---

[64] Timothy C. Lehmann, "Keeping Friends Close and Enemies Closer: Classical Realist Statecraft and Economic Exchange in U.S. Interwar Strategy." *Security Studies* 18, no. 1 (2009): 115–47. For a detailed analysis of the adaptive mechanisms that lessen the dependency-strategic concerns related to oil for American national security, see Eugene Gholz and Daryl G. Press, "Protecting 'The Prize': Oil and the U.S. National Interest," *Security Studies* 19, no. 3 (2010): 453–85.
[65] Roy Licklider, "The Power of Oil: The Arab Oil Weapon and the Netherlands, the United Kingdom, Canada, Japan, and the United States," *International Studies Quarterly* 32, no. 2 (1988): 205-226.
[66] Hirschman 1945, 18-19.

Strategic assets can be characterized by multiple complementary logics (Figure 1). Oil, for instance, is the prototypical strategic asset because it activates each of these strategic logics in a powerful way. Oil is *cumulative-strategic*, in that it gives rise to a valuable industry characterized by an oligopolistic global market structure. It is *infrastructure-strategic*: its integration leads to a broad upgrading of economic and military systems, for which no sub-state actor has full incentives to adequately provide. It is *dependency-strategic*: it serves as a critical flow input which is vulnerable to being cut-off.

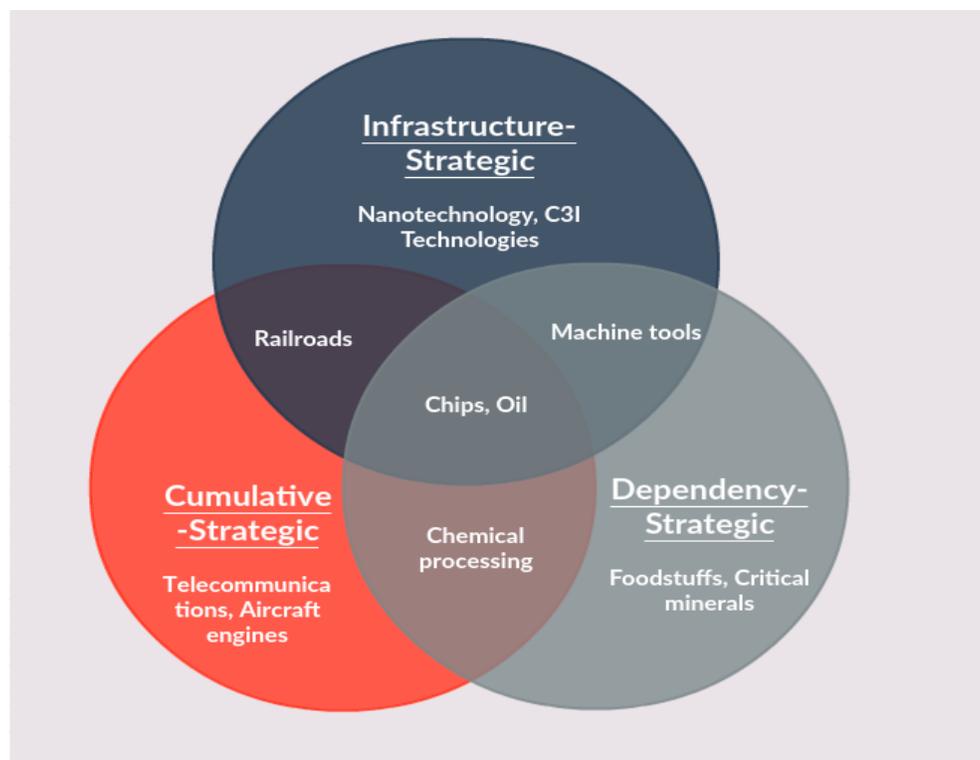

*Figure 1*

Chips are another asset that has exhibited all three of our framework's logics.[67] For large portions of its history, the semiconductor industry has been defined by an oligopoly structure. Only a few firms are able to invest the high capital and R&D expenditures and accumulate the experience required to keep up with constant technical iterations. In addition, investments in integrated circuit development feed into advances in computers, machine tools, and robots, thereby generating diffuse productivity across the entire electronics value chain.[68] Finally, chips are dependency-strategic in both domains. Only a small group of foundries can design and/or fabricate the microchips that power a range of crucial military platforms, including aircraft, electronic warfare systems, and radar. These foundries also make the chips that are a critical input across a wide range of information industries.[69]

---

[67] Chips refers to integrated circuits, which constitute a large segment of the semiconductor industry.

A second type of relationship arises when the logics tradeoff with each other. States seeking to capture a positive externality from one strategic logic may expose themselves to a negative externality from another strategic logic. Churchill encountered such a tradeoff between the infrastructure-strategic and dependency-strategic logic when deciding whether to convert the British navy to oil-burning ships. On the one hand, oil-powered ships presented a significant upgrade in operational efficiency, speed, and radius of action, which would provide a spillover boost to the range and capabilities of the entire military. On the other hand, the transition away from ships powered by steam coal, which was abundant in British mines, would make the navy reliant on oil imports from distant countries.[70]

These tradeoffs underscore that goods and technologies can be strategic in more than one sense. While the existing literature on the economics of national security has examined these logics separately, we package them together under a comprehensive framework that can inform a state's overall technology strategy. The following case study of U.S.-Japan technological rivalry, and the application to current strategizing about artificial intelligence, demonstrate the value of our approach.

---

[68] Crawford, "Economic Vulnerability in International Relations, 53.
[69] Masaru Yoshitomi, "New Trends of Oligopolistic Competition in the Globalisation of High-Tech Industries: Interactions Among Trade, Investment, and Government," in *Strategic Industries in a Global Economy*, ed. OECD, 15-35, 19.
[70] Yergin, *The Prize*, 156.

In advance of the empirical analysis, it is important to clarify how the case study supports our framework. Our diagnostic theory for strategic assets uncovers logical inconsistencies in the end-means chain by which national strategists target specific technologies (means) to accumulate power and plenty (end). In principle, our theory does make positive causal claims that can be tested, given sufficient auxiliary assumptions: if a state better identifies strategic assets, then all else equal it will tend to better accrue wealth and power. However, similar to deterrence theory, which George and Smoke state is "best understood as a contingent policy theory," these claims cannot be fully tested due to many confounders.[71] Technology policy is one of many instruments (i.e. many end-means chains), and strategic asset identification is only one step in the implementation of technology policy (i.e. many other steps in the end-means chain). Thus, the role of empirics in this work is not to conduct systematic evaluation of causal relationships, but rather to illustrate the framework's utility for improving the identification of strategic assets.[72]

---

[71] George and Smoke, "Deterrence and Foreign Policy," 181
[72] George and Smoke, *Deterrence in American Foreign Policy*, 622.

## Part III: Strategic Assets in U.S.-Japan Technological Rivalry

In the 1980s and 1990s, the U.S. confronted a changing international landscape of power. As the Soviet Union neared its fall, Japan emerged as a challenger to U.S. technological preeminence, sparking concerns over the U.S. ability to remain a leader in critical fields.[73] Alarmed by Japanese technological ascendancy, the U.S. government, independent academics, and industry associations published dozens of major lists of critical technologies. Labeled as a "critical technologies movement," these efforts aimed to identify strategic assets in a systematic and comprehensive manner.[74]

Our case study focuses on the rationale behind technology assessments in the U.S. during the 1980s and 1990s for a variety of reasons. First, as evidenced by the critical technologies movement, there is a rich literature and empirical record to pore through. Compared to other potential cases, U.S.-Japan competition generated a disproportionately large amount of deliberation over strategic goods and technologies, making it a crucial test of our framework's analytical value.

In addition, commentators note the parallels from this case to the current period. China's rising challenge to American technological dominance has raised similar concerns in U.S. policymaking circles about protecting strategic assets.[75] Thus, the case is substantively important for strategy related to U.S.-Sino competition. Lastly, it is important to select a case where enough time has passed so as to better assess the wisdom of labeling certain assets as strategic. The U.S.-Japan case fits the bill.

---

[73] From a list of twenty technologies, the DoD's critical technologies plan identified five technology groups critical to military superiority in which Japanese firms had the lead. U.S. Department of Defense, *Critical Technologies Plan* (Washington, DC: Report to the Committee on Armed Services, U.S. Congress, 1990).

[74] Mary Ellen Mogee, *Technology Policy and Critical Technologies: a Summary of Recent Reports* (Washington, DC: The National Academies Press, 1991), 24; Caroline S. Wagner and Steven W. Popper, "Identifying Critical Technologies in the United States: A Review of the Federal Effort," *Journal of Forecasting* 22, no. 2–3 (2003): 113–28.

[75] Stephen S. Roach, "Japan Then, China Now." *Project Syndicate*, 27 May 2019, https://www.project-syndicate.org/commentary/for-america-china-is-the-new-japan-by-stephen-s-roach-2019-05. We also considered the U.S. response to the Sputnik in the late 1950s, which also generated much debate over which technologies contribute disproportionately to national security. Ultimately, we decided that the dynamics of the U.S.-Japan case, especially the enmeshing of economic competitiveness and national security concerns, were a better test of our framework's applicability to the current period.

To preview the results, the case study evidence confirms the analytical value of our conceptual framework. By feeding into ineffective industrial policy and flawed assessments of relative technological capabilities, mis-identification of strategic assets hampered U.S. economic and military competitiveness in this period. We also present evidence that the underlying logics of our framework informed U.S. policymaking in isolated cases. However, the critical technologies movement's effectiveness was hindered by its failure to integrate all three logics. These missteps, if our framework has diagnostic value, should consistently involve the identification of an asset as "strategic" which our framework would have excluded (*false positives*) and the failure to identify an asset as "strategic" which our framework would have included (*false negatives*).

*False Positives and Negatives*

To demonstrate the utility of our theoretical framework, we highlight examples of *false positives* and *false negatives* in the assessment of strategic assets, for which our three-logic framework could have improved technology strategy. Regarding false positives, U.S. thinkers used the "high-tech" designation as a blanket label for strategic industries, a fixation that Ostry and Nelson describe as "high-tech fetishism."[76] While some of the "high-tech" sectors could be justified as cumulative-strategic (e.g. supercomputers, discussed below), the "high-tech" designation was overly broad and produced flawed indicators of U.S. competitiveness. For example, a 1986 report by the Department of Commerce (DOC) claimed that the U.S. had a $2.6 billion trade deficit in high-tech industries, resulting in the creation of a potpourri of councils, commissions, and institutes to study all the

---

[76] Sylvia Ostry and Richard Nelson, *Techno-Nationalism and Techno-Globalism* (Washington, DC: Brookings Institution Press, 1995), 60.

varieties of strategic technologies.[77] But when this high-tech trade deficit was measured in high-tech products rather than high-tech industries, a deficit of $17 billion (1985-1988) turned into a surplus of $3.5 billion. In this case, analysts had included trade in scales, cash registers, and similar products (which did not generate cumulative-strategic externalities) in the indicator deficit because they fell under the DOC's "Office and Computing Machines" high-tech industry classification.[78]

In their attempts to apply the infrastructure-strategic logic, analysts treated biotechnology as an industry rather than a set of techniques that affect a wide range of industries. Four influential technology reports identified biotechnology — largely framed toward biomedical applications —as a critical technology.[79] However, none of these reports assessed whether existing federal funding for biotechnology was sufficient. In fact, a separate OTA assessment recommended that, given the success of existing private efforts, Congress consider reducing federal funds for basic biomedical research and redirecting them to biotech applications in other sectors such as agriculture, chemicals, and waste management.[80] These recommendations underscore the need to clearly specify the contexts in which a set of techniques qualifies as infrastructure-strategic.

Finally, U.S. policymakers misdiagnosed the dependency-strategic logic in numerically controlled machine tools, marking a failure to zoom into the key technologies of a general sector as well as gauge the extent to which supply was concentrated in a particular country. In 1986, the U.S. government designated the entire industry of machine tools as strategic, negotiating a Voluntary Restraint Agreement (VRA) with Japan and other countries to limit the level of their machine tool exports to the U.S. for a period of five years. The implementation of VRAs was relatively insulated from factors unrelated to the strategic assets framework, such as lobbying by the domestic machine tool firms or labor conditions in the machine tool industry.[81] Rather, national security concerns, related to the importance of machine tools for the U.S. defense base, were the main driver of protectionism.

---

[77] Benoit Godin, "The Obsession for Competitiveness and its Impact on Statistics: the Construction of High-technology Indicators," *Research Policy* 33, no. 8 (2004): 1217-1229.
[78] Thomas A. Abbott III, "Measuring High Technology Trade: Contrasting International Trade Administration and Bureau of Census Methodologies and Results," *Journal of Economic and Social Measurement* 17, no. 1 (1991): 17-44.
[79] Mogee, "Technology Policy and Critical Technologies," 25-29.
[80] Office of Technology Assessment, *Biotechnology in a Global Economy* (Washington, DC: U.S. Government Printing Office, 1991).

A fuller appreciation of dependency concerns would have better qualified which machine tool assets were most strategic for enhancing U.S. national security. At the time, the general category of machine tools also included mature, standardized product models for which production was not concentrated in any particular country, limiting the salience of the dependency-strategic logic. In fact, new suppliers in Belgium, Denmark, Italy, and Spain were beginning to further reduce the U.S.'s dependence on Japan in some types of machine tools. Counterproductively, the VRA restricted the expansion of these new suppliers. Instead, analysts should have focused their attention on specific sub-sectors of machine tools, such as grinders of ceramic and other non-metallic materials, where Japan dominated U.S. imports of these key sub-sectors.[82]

Because of the tendency of interest groups and policymakers to deploy the "strategic" descriptor broadly, uncovering the existence of false positives represents a relatively easy test of the theory. We also find evidence of false negatives, which constitute a harder test. For instance, military planners neglected dependency-strategic risks associated with rayon fibers. In November 1988, the American apparel company Avtex announced that it was closing down due to foreign competition, sending shockwaves through the U.S. military and space community, as Avtex was the only producer of rayon fibers critical to the production of missiles and rockets.[83] Alternative sources could have been certified and other fibers could have been adapted into substitutes, but this process would have taken longer than the period of time the available supply of rayon would support production.

---

[81] The U.S. machine tool industry was neither highly concentrated nor composed of a small number of firms, which reduced its capacity to organize and lobby for protection. In addition, it was relatively easy for machine-tool builders to find jobs in other industries, which means that adjustment assistance concerns cannot explain protectionist actions. For a systematic analysis of the causes of VRAs in machine tools, see Elias Dinopoulos and Mordechai E. Kreinin, "The US VER on Machine Tools: Causes and Effects," in *Empirical Studies of Commercial Policy*, ed. Robert E. Baldwin (Chicago: University of Chicago Press, 113-134), 114-121.
[82] Theodore H. Moran, "The Globalization of America's Defense Industries: Managing the Threat of Foreign Dependence," *International Security* 15, no. 1(1990): 57-99, 87-88.

Though the U.S. government and aerospace industry officials eventually negotiated a deal to keep Avtex open, the case illustrates that the defense community undervalued risks with relying on sole-source supplies for key inputs.[84] Considerations of military utility dominated discussions of strategic assets. The DoD's annual critical technology plans defined the criticality of technologies based on their importance to U.S. weapons systems' long-term qualitative superiority.[85] Under this framework, cases like Avtex fell through the cracks because, as the U.S. Defense Science Board concluded, "neither DoD nor industry ha[d] the means of measuring the scope of [foreign] dependence or of identifying the systems and components which are affected."[86]

The U.S. military also overlooked infrastructure-strategic aspects of software technologies. There was no question that both the U.S. and Japan recognized the growing *importance* of software for national security. In 1990 the DoD identified "software producibility" as one of 20 technologies "most essential" to the long-term superiority of U.S. weapon systems.[87] Software codes were also at the heart of a 1989 controversy over the U.S.-Japanese co-development of the "FSX" fighter aircraft.[88]

---

[83] Office of Technology Assessment, *Holding the Edge: Maintaining the Defense Technology Base* (Washington, DC: U.S. Government Printing Office, 1989), 33.
[84] A concentration of foreign suppliers in a particular asset is not necessary for the dependency-strategic logic to be in play. For instance, relying on a sole-source domestic supplier could also present a negative externality if that source was vulnerable to natural disasters, cyberattacks, and other disruptions.
[85] Crawford, "Economic Vulnerability in International Relations," 16-17.
[86] Defense Science Board, "Summer Study on The Defense Industrial and Technology Base" (Office of the Under Secretary of Defense for Acquisition, Washington, DC, 1988), 51.
[87] U.S. Department of Defense, *Critical Technologies Plan*, A-30-A-32.
[88] Michael Mastanduno, "Do Relative Gains Matter? America's Response to Japanese Industrial Policy," *International Security* 16, no. 1 (1991): 73–113, 84-83

It was the infrastructure-strategic characteristics of software that deserved more attention. By the mid-1980s, escalating software costs and uncoordinated software development — 400 different programming languages and variations were used across the DoD's weapons systems — led the U.S. military to acknowledge it was facing a "software crisis."[89] Across major weapon systems, software development issues contributed to cost overruns, fielding delays, and even entire program cancellations.[90] In response, the DoD launched a Software Initiative in an effort to create a central software authority and promote a standard programming language called ADA.[91] But the budget of the office in charge of integrating ADA across the Pentagon peaked at $7 million, and "software R&D was the first to go" amidst defense cuts during the late 1980s.[92]

The DoD neglected the infrastructural aspects of software development regarding two undertakings in particular: taking advantage of commercial software spillovers and overcoming individual military services' autonomy over their software projects. Regarding the first factor, the Pentagon's ADA initiative, though attuned to infrastructure-strategic concerns in the military domain, actually limited convergence with U.S. commercial software applications, which were advancing much more quickly. One OSD executive admitted that "the Department would have improved cost and quality performance more by standardizing on any existing programming language instead of creating Ada."[93] A 1989 OTA report concluded that divergence between government and commercial software development had produced "separate defense and commercial businesses that often do not share technology."[94]

---

[89] Goldstein, "Institutional Resistance to the Demands of New Information Technology," 30; Office of Technology Assessment, *Holding the Edge*, 170.
[90] Henry Attanasio, "Contracting for embedded computer software within the Department of the Navy" (Master's thesis, Naval Postgraduate School, 1990), 11-13.
[91] Goldstein, "Institutional Resistance to the Demands of New Information Technology," 30-32.
[92] Ibid., 32.
[93] Ibid., 40. Technically savvy officers would program in the C++ language instead of ADA (Lindsay 2010, 632).
[94] Office of Technology Assessment, *Holding the Edge*, 35.

As for the second factor, without special attention to the needs for major organizational reforms in order to adapt software across the military, individual services "ferociously resisted" the standardization of software development practices.[95] When the Office of the Secretary of Defense (OSD) convened defense contractors to enforce cross-service software standardization, the U.S. Air Force sent no representatives to the conference, essentially removing the development of the advanced tactical fighter, one of the most complex software systems to date, from the standardization effort.[96]

*The Logics in Isolation*

In many instances, U.S. policymakers and analysts reasoned in line with our three logics, with varying degrees of clarity and explicitness. Based on amplified cumulative gains in supercomputers compared to steel, U.S. policymakers more actively protected the former. Despite their small market relative to steel, the production of supercomputers was highly dependent on "learning by doing" in which experience developing previous generations transferred to developing the next generation, and early market share enabled the development of unique libraries of software..[97] In contrast, steel production technology was more easily diffusible as the technology had become standardized,[98] which was partly why the U.S. was more willing to make concessions on trade disputes over steel.[99]

U.S. policymakers also appropriately attended to the infrastructure-strategic logic in some cases. An influential Council on Competitiveness report published during this time period, titled

---

[95] Goldstein, "Institutional Resistance to the Demands of New Information Technology," 36.
[96] Ibid., 36. The advanced tactical fighter was one of the five largest military software systems in 1990. U.S. Department of Defense, *Critical Technologies Plan*, A-26.
[97] By 1993, American firms were able to draw on these cumulative gains to dominate most markets, including 85 percent of the European public sector market — though they failed to penetrate Japan, which realized its industry could not compete without substantial protection. John C. Matthews III, "Current Gains and Future Outcomes: When Cumulative Relative Gains Matter," *International Security* 21, no. 1 (1996): 112-146, 130-134.
[98] Generally, cumulative advantages fade as a technology matures, design parameters become standardized, and greater competition emerges through channels such as incremental refinement, distribution, and marketing. One exception is integrated circuits, for instance, where the number of transistors on a chip has doubled every two years while the costs are halved ("Moore's Law"). Borrus et al. 1986, "Creating Advantage," 104.
[99] Matthews III, "Current Gains and Future Outcomes," 140-142.

"Gaining New Ground: Technology Priorities for America's Future," emphasized support for "critical generic technologies" that could potentially enable growth across a range of industries.[100] Of these, microelectronics were a prime target for technology policy such as the Strategic Computing Initiative. This was justified because advances in microelectronics provided "infrastructural support for all computer development."[101] In a widely-read text, Laura Tyson, who later served as Chair of President Clinton's Council of Economic Advisers, articulated a "cautious activist" policy toward industries in which "the returns to technological advance create beneficial spillovers for other economic activities, and barriers to entry generate market structures rife with first-mover advantages and strategic behavior."[102] This reflects the infrastructure- and cumulative-strategic logic, respectively.

U.S. technology policy was most attuned to dependency concerns. During this period, the U.S. government produced a total of sixteen different studies that assessed the globalization of U.S. defense production.[103] Defense industrialists pinpointed America's increased reliance on foreign suppliers for key components of weapon systems. For instance, Japanese companies such as NEC and Mitsubishi dominated the production of gallium arsenide (GaAs), a key material used in field effect transistors that enabled higher computing speeds and radiation resistance for missile guidance and radar.[104]

Addressing the vulnerability to supply cut-offs was not straightforward. With a defense technology base that increasingly relied on globalized, dual-use industries, the military faced a tradeoff between taking advantage of these infrastructure-strategic dual-use assets, or "going it alone" to avoid dependency-strategic risks. Indeed, the technologies where foreign dependence was most pronounced — e.g. advanced semiconductors, structural materials, and fiber optics — were also those driving innovation and qualitative improvements in critical military systems.[105]

---

[100] Council on Competitiveness, *Gaining New Ground: Technology Priorities for America's Future* (Washington, DC: Council on Competitiveness, 1991).
[101] Alex Roland and Philip Shiman, *Strategic Computing: DARPA and the Quest for Machine Intelligence, 1983-1993* (Cambridge, MA: MIT Press, 2002), 33.
[102] Laura D. Tyson, *Who's Bashing Whom?* (Washington, DC: Peterson Institute for International Economics, 1993), 3.
[103] Stephen G. Brooks, "Reflections on Producing Security," *Security Studies* 16, no. 4(2007): 637-678, 667.
[104] Seventy-five percent of the GaAs material for field effect transistors were obtained from foreign sources, mainly Japanese companies. National Research Council, *Foreign Production of Electronic Components and Army Systems Vulnerabilities* (Washington, DC: The National Academies Press, 1985), 26.

While the above examples illustrate how isolated aspects of U.S. technology strategy did heed the three logics, the logics were not integrated into identifying strategic technologies. One systematic review of six reports on critical technologies, all of which were published between 1987-1991, found that the lists relied on broad definitions of technology and different criteria, time horizons, and methodologies.[106] The National Critical Technologies Panel — established in 1990 as the principal instrument for the federal government to answer the question "what is a critical technology" — was the source of one of these assessments, but it gave no criteria for determining what is "critical," thereby leaving the matter in the hands of 13 individuals from government and industry."[107] This was the case for other efforts to identify critical technologies. "Most of the reports involve little or no serious original research or data collection and little or no guiding theoretical framework," writes Mogee."[108]

Absent a guiding theoretical framework, efforts to identify strategic assets fell prey to familiar traps. The principal purpose of exercises to name key technologies turned toward evaluations of an asset's absolute utility (*importance*) for U.S. national security and economic prosperity rather than the level of intervention an asset demands from the federal government (*externality*). Wagner and Popper conclude that the "critical technologies reports must be held to have had little formal effect on US federal policy towards technology development."[109] Of course, it has to be recognized that there were many other factors at play, including the U.S.'s decentralized technology system and its ideological orientation against picking technology winners. But the evidence above does point to confusion over how to define critical technologies as a key part of the explanation for the limited usefulness of critical technology identification in this period.

---

[105] At the time, the Office of Technology Assessment issued a series of annual reports that focused on this tradeoff: Office of Technology Assessment, *Holding the Edge*; Office of Technology Assessment, *Arming our Allies*; Office of Technology Assessment, *Redesigning Defense*.
[106] Mogee, "Technology Policy and Critical Technologies," 37-38.
[107] Wagner and Popper, "Identifying Critical Technologies in the United States," 117-118.
[108] Mogee, "Technology Policy and Critical Technologies," 37.

One last note about how the above case should be interpreted. Our framework elaborates three logics that can make an asset worth attending to. Our theoretical claim is that when these logics are operative, states who attend to the assets and adopt appropriate policies will gain in power and plenty. Ideally, we could construct a counterfactual for the U.S.-Japan case in which decision-makers fully adopted our view of strategic assets, and thereby isolate the effects on U.S. economic and military might. One could imagine a world where, for instance, the U.S. military appropriately identified the infrastructure-strategic aspects of software advances in the commercial realm and standardized its programming language accordingly. This preventative approach could have mitigated software bottlenecks, which continue to plague military systems decades later.

Given the complexities of the end-means chain from strategic asset identification to end goals, the above empirics are not intended to prove the theoretical claim. A counterfactual in which the U.S. addressed dependency-strategic concerns in machine tools, along the lines of our framework, must also grapple with other factors that outweigh the benefits of improved strategic asset identification. For instance, Japan's macroeconomic struggles hampered its machine tool industry, which nullified the long-term vulnerability of the U.S. in this domain. Still, our findings do illustrate the importance of strategic asset identification across a range of domains and scenarios, and they show how conceptual clarity could have improved policy around strategic assets.

---

[109] Wagner and Popper, "Identifying Critical Technologies in the United States," 123.

## Part IV: Implications

How should strategists identify the strategic assets of the current era? We conclude by applying our theoretical framework to the contemporary case of artificial intelligence (AI), a technology that has drawn so much high-level attention from states that the World Economic Forum (WEF) has published a framework to help governments create "minimum viable" national AI strategies.[110] The WEF framework describes AI as the "engine that drives the Fourth Industrial Revolution."[111] Others note AI's potential to transform the military balance of power.[112] In sum, the governance of advanced AI systems "may be the most important global issue of the 21st century."[113]

*Strategic Assets in the AI Era*

While most agree on AI's importance, there is less clarity over the strategic aspects of developments in AI. Though all three strategic logics are present with AI, and in some cases trade off against each other, infrastructural considerations are most central to what makes AI strategic. Described by leading economists as a new general-purpose technology (GPT), advances in AI can potentially transform a wide range of economic sectors.[114] If the trajectory of previous GPTs holds, the effective diffusion of AI will take decades and many complementary innovations.[115] For example, the introduction of the steam turbine, which occurred about a decade after the invention of the dynamo, was crucial to the spread of electric power across manufacturing industries.[116]

---

[110] World Economic Forum, "A Framework for Developing a National Artificial Intelligence Strategy," 4 October 2019, https://www.weforum.org/whitepapers/a-framework-for-developing-a-national-artificial-intelligence-strategy
[111] Ibid., 4.
[112] Kenneth Payne, "Artificial Intelligence: A Revolution in Strategic Affairs?" *Survival* 60, no. 5 (2018): 7–32; Michael C. Horowitz, "Artificial Intelligence, International Competition, and the Balance of Power," *Texas National Security Review* (2018).
[113] Allan Dafoe, "AI Governance: A Research Agenda," Governance of AI Program, 27 August 2018, https://www.fhi.ox.ac.uk/wp-content/uploads/GovAI-Agenda.pdf, 5.
[114] Brynjolfsson, Erik, Daniel Rock, Chad Syverson. 2018. The Productivity J-Curve: How Intangibles Complement General Purpose Technologies. NBER Working Paper 25148; Cockburn, Iain M, Rebecca Henderson, and Scott Stern. "The Impact of Artificial Intelligence on Innovation." National Bureau of Economic Research Working Paper. March 2018.
[115] Brynjolfsson et al, "The Productivity J-Curve"; Thurbon and Weiss, "Economic Statecraft at the Frontier," 18.
[116] Vaclav Smil, *Creating the Twentieth Century* (Oxford, UK: Oxford University Press, 2005), 33-97.

The breadth of spillovers and prolonged payoff period associated with AI makes it difficult for private investors to capture most of the gains. One study of the evolution of six GPTs found that large-scale, long-term government investment was necessary in accelerating their commercial development.[117] Thus, recognizing the strategic implications of AI as a GPT is the first step to realizing its potential for economic transformation.

The general-purpose nature of AI should also guide defense industrial policy. If AI is "much more akin to the internal combustion engine or electricity than a weapon," as Horowitz argues, then military strategists should pay more attention to the organizational adaptations that advances in AI may demand.[118] A recent study warns that the DoD "has not yet adapted its enterprise processes to effectively support the rapid and widespread adoption warranted by the potential benefits (of autonomous capabilities)."[119]

These organizational challenges have been raised by previous GPTs. After all, AI systems are still built on software, so many of the lessons from the U.S. military's attempt at software standardization in the 1980s still hold. According to a recent submission to the National Security Commission on Artificial Intelligence's (NSCAI) call for ideas, military AI projects are isolated from best practices in the civilian sector. Either the DoD adapts its software ecosystem to better cross-pollinate with the private sector's software ecosystem or it "risk[s] losing out to China and Russia."[120]

---

[117] Vernon W. Ruttan, *Is War Necessary? Is War Necessary for Economic Growth?* (Oxford, UK: Oxford University Press, 2006).
[118] Michael C. Horowitz, "The Algorithms of August," *Foreign Policy*, 12 September 2018, https://foreignpolicy.com/2018/09/12/will-the-united-states-lose-the-artificial-intelligence-arms-race/
[119] Defense Science Board, "Report of the Defense Science Board Summer Study on Autonomy" (Office of the Under Secretary of Defense for Acquisition, Technology, and Logistics, Washington, DC, 2016), 98.
[120] James Ryseff, "How to (Actually) Recruit Talent for the AI Challenge," *War on the Rocks*, 5 February 2020, https://warontherocks.com/2020/02/how-to-actually-recruit-talent-for-the-ai-challenge/

The interaction of AI with all three logics is partly what makes strategic thinking about the technology so difficult, so strategists should not neglect the other two logics. Regarding the cumulative-strategic considerations in the economic domain, big data platforms, such as Facebook and Amazon, appear to benefit from a virtuous circle that links access to data, improvements in machine learning models, and the attraction of more users and data. Some regulators, who believe that these platforms are exploiting this market power, are exploring options to restore competition in this domain.[121] As for military applications, cumulative-strategic dynamics vary by application — a nuance lost in the narrative of an AI arms race.[122] It may be much harder for advanced militaries to sustain large first-mover advantages in military applications of AI which build directly off of open research in the civilian sector, such as image recognition for reconnaissance and predictive analytics for logistical planning.[123] For advanced weapon systems for which autonomous capabilities demand the integration of AI into more complex systems, the cumulative-strategic dynamics may be much more salient.[124]

Of the three logics, the dependency-strategic logic has drawn a disproportionate share of the attention when it comes to AI. This may be partly due to the overwhelming focus of the existing statecraft literature on tools such as trade and financial sanctions.[125] In particular, analysts and policymakers have identified AI hardware as a strategic asset for U.S.-China technological competition. As Tim Hwang writes, "The extent to which the U.S. is able to successfully deny China access to advanced computing power, and the extent to which China is able to develop it domestically or acquire it otherwise, remains to be seen."[126]

---

[121] Jerrold Nadler and David N. Cicilline, "Investigation of Competition in Digital Markets" (Majority Staff Subcommittee on Antitrust, Commercial and Administrative Law, 2020); Steven Weber, "Data, Development, and Growth," *Business and Politics* 19, no. 3 (2017): 397–423.
[122] Zwetsloot, Remco, Helen Toner, and Jeffrey Ding, "Beyond the AI Arms Race," *Foreign Affairs*, 16 November 2018, https://www.foreignaffairs.com/reviews/review-essay/2018-11-16/beyond-ai-arms-race
[123] Horowitz, "Algorithms of August."
[124] Gilli and Gilli, "Why China Has Not Caught Up Yet," 189.
[125] Thurbon and Weiss, "Economic Statecraft at the Frontier," 5.

Statecraft targeted at the dependency-strategic aspects of the AI supply chain is complicated by tradeoffs between various logics. Take, for example, U.S. policy debates over the strategic asset of semiconductor manufacturing equipment (SME), an integral piece of the supply chain in hardware for training and execution of AI algorithms. One of the four initial consensus judgements in the NSCAI's first report was that the U.S. government should continue to use export controls to protect American advantages in AI hardware, in particular those in semiconductor manufacturing equipment.[127]

On the one hand, export controls would leverage the dependency-strategic aspects of SME, since U.S. firms occupy around 50 percent of the global market. However, the SME industry is also strategic in the cumulative sense, for which the optimal policy involves maximizing global revenues. Thus, exploiting the dependency-strategic externalities of SME to gain leverage over China (e.g. restricting China's military advances) could enable cumulative-strategic gains for U.S. competitors, as SME vendors in Europe and Japan could benefit from the loss of U.S. SME sales to China.[128]

*Conclusion*

As revealed by our preliminary analysis of the strategic aspects of AI, our framework aids strategic asset identification by assessing how the characteristics of a particular technology interact with the surrounding context. Most studies of economic statecraft and defense industrial policy concentrate on relevant actors, strategies, and consequences. Significantly less emphasis is placed on the objects themselves — the strategic assets which are often the target of military competition and foreign economic policy. The fundamental contribution of this article is toward resolving a tension in the international landscape today: even as nations are increasingly concerned about building up their advantage in strategic goods and technologies, much more work needs to be done to understand the underlying logic of what makes an asset strategic.

---

[126] Tim Hwang, "Computational Power and the Social Impact of Artificial Intelligence," 23 March 2018, https://papers.ssrn.com/sol3/papers.cfm?abstract_id=3147971
[127] National Security Commission on Artificial Intelligence, "Interim Report" (National Security Commission on Artificial Intelligence, 2019).
[128] John Verwey, "The Health and Competitiveness of the U.S. Semiconductor Manufacturing Equipment Industry," (Office of Industries Working Paper ID-058, 2019), 19; Jade Leung, Sophie-Charlotte Fischer, and Allan Dafoe, "Export Controls in the Age of AI," *War on the Rocks*, 28 August 2019, https://warontherocks.com/2019/08/export-controls-in-the-age-of-ai

Future research could expand the scope of our framework. A continuing issue with determining strategic technologies is the level of aggregation. Does "strategic" apply to specific technological innovations, classes of techniques, technological systems, or entire sectors? Theoretically, strategic assets could even encompass things other than goods and technologies. For example, governments compete to attract highly-skilled talent in science and technology in what the International Organization for Migration has labeled a "human capital accretion 'sweepstakes.'"[129]

Moreover, our theory can encompass more than just three strategic logics. Consider a brief sketch of the "poisoned-chalice logic," which highlights the externalities associated with hardware hacks from upstream parts and components. While closely related to risk from supply disruptions, the poisoned chalice refers to an adversary's access to the asset in an upstream portion of the supply chain. For instance, datasets can be "poisoned" to attack the integrity of the AI systems that are trained on them, such that an adjustment to just a single observation can produce a "backdoor" that can later be exploited.[130] If the complexity of supply chains and the number of suppliers for many advanced technology systems continues to increase, this externality will grow in relevance.[131]

---

[129] High-skilled migrants provide "free" knowledge assets to receiving states since sending states bore the costs of education and training. This relates to our framework, as global talent flows can be considered a positive externality for the receiving state. Fiona B Adamson, "Crossing borders: International migration and national security," *International Security* 31, no. 1 (2006): 165-199, 186..

[130] Ali Shafahi, W. Ronny Huang, Mahyar Najibi, Octavian Suciu, Christoph Studer, Tudor Dumitras, and Tom Goldstein, "Poison Frogs! Targeted Clean-label Poisoning Attacks on Neural Networks," in *Advances in Neural Information Processing Systems* (2018), 6103-6113.

[131] For studies of risks to the DoD's supply chain related to this logic, see: Brooks, "Reflections on Producing Security," 671-672.

While many extensions are possible, one consistent element of our framework is that mapping the logic behind strategic goods and technologies is only a starting point. It would be a mistake to leap from the identification of a strategic asset to the implementation of industrial policy targeted at that asset. There are many reasons to be skeptical of technological planning, forecasting, and selection by central authority. Cost overruns, wasteful rent seeking, and crowd-out from "picking winners" all contribute to "government failure," which could outweigh the benefits of correcting market failures.[132]

Still, our effort to bring conceptual rigor to the discussion of strategic assets is a prerequisite to effective strategy. One assessment of national efforts to enhance scientific and technological competitiveness, which synthesized over fifty years of research on national innovation rates, found that the one common trait among the successful countries is "their dedication, not to particular institutions or policy designs, but to *solving market failures and network failures in general.*"[133] In essence, our paper translates this insight into a framework for the identification of strategic assets.

Fred Halliday called international relations the capstone discipline of the social sciences in part because it is tasked with integrating concepts from all the other sciences. An economist may identify strategic assets as the civilian technologies of greatest economic importance; a military planner may think of strategic assets as those that are most essential to military operations; a historian may understand strategic assets as those that have had the most significant effects in shaping the development of society. Our paper offers a framework for how a grand strategist should conceive of strategic assets.

---

[132] Kirshner, "Political Economy in Security Studies after the Cold War," 77.
[133] Taylor, *The Politics of Innovation*, 277. Emphasis ours.